\documentclass[12pt]{article}
\pdfoutput=1

\usepackage{graphicx,array}
\usepackage{color}
\usepackage{latexsym}
\usepackage{amsthm}
\usepackage{amsmath}
\usepackage{amssymb}
\usepackage[numbers,sort&compress]{natbib}
\usepackage{bm}
\usepackage{slashed} 
\usepackage{mathrsfs}
\usepackage{hyperref} 
\hypersetup{
    colorlinks=true,       
    linkcolor=red,          
    citecolor=blue,        
    filecolor=magenta,      
    urlcolor=blue           
}
\usepackage[all]{hypcap} 

\usepackage[font={small}]{caption}

\pdfoutput=1
\usepackage{graphicx}
\usepackage{epstopdf}
\usepackage{dcolumn}
\usepackage{bm}
\usepackage{hyperref}
\usepackage{color}
\usepackage{amsmath}
\usepackage{mathrsfs}
\usepackage{cancel}
\usepackage{xpatch}

\newcommand{\be}{\begin{equation}}
\newcommand{\ee}{\end{equation}}

\newcommand{\GeV}{{\rm GeV}}
\newcommand{\TeV}{{\rm TeV}}

\begin{document}
\thispagestyle{empty}

\begin{flushright}
CERN-TH-2019-041
\end{flushright}
\bigskip 
\begin{center}
{\Large \bf Quark masses, CKM angles and \\ Lepton   Flavour Universality violation } \\
\vspace{1cm}
{\large Riccardo Barbieri$^a$ and Robert Ziegler$^b$} \\
\vspace{.4cm}
{$^a$\it \small Scuola Normale Superiore, Piazza dei Cavalieri 7, 56126 Pisa, Italy and INFN, Pisa, Italy} \\
{$^b$\it \small CERN, Theoretical Physics Department, Geneva, Switzerland }
\vspace{1cm}
\end{center}

\begin{abstract}\noindent
A properly defined and suitably broken $U(2)$ flavour symmetry leads to successful quantitative relations between quark mass ratios and CKM angles. At the same time the intrinsic distinction introduced by $U(2)$ between the third and the first two families of quarks and leptons may support anomalies in charged and neutral current semi-leptonic $B$-decays of the kind tentatively observed in current flavour experiments. We show how this is possible by the exchange of the $(3,1)_{2/3}$ vector leptoquark  in two $U(2)$-models with significantly different values of Lepton Flavour Universality violation, observable in foreseen experiments.  \\
\end{abstract}

\vfill
\noindent\line(1,0){188}
{\scriptsize{ \\ \texttt{$^a$ \href{mailto:riccardo.barbieri@sns.it}{riccardo.barbieri@sns.it}\\ $^b$ \href{mailto:robert.ziegler@cern.ch}{robert.ziegler@cern.ch}}}}

\newpage

\section{Introduction and statement of the framework}
\label{intro}

In spite of several attempts, a truly convincing way of reducing the number of free parameters in the flavour sector of the Standard Model is still elusive. To the point that one can express a pessimistic view about making progress in this area without  new crucial experimental information. In this respect, the apparent presence of Lepton Flavour Universality (LFU) violations in B-decays represents an interesting possibility that we want to explore in this article.
As observed in previous works~\cite{Barbieri:2015yvd,Barbieri:2016las,Buttazzo:2017ixm}, a putative anomaly in the decays of a third generation particle~\cite{Lees:2013uzd,Aaij:2015yra,Hirose:2016wfn,Aaij:2014ora,Aaij:2017deq,Aaij:2017vbb} invites to make a connection with the relative separation between the third and the first two generations, both as to their masses and to the CKM angles. In turn this may call into play a $U(2)$-symmetry that acts on the first two generations as doublets and the third generation particles as singlets.

As recalled in Section~\ref{existenceproof}, a properly defined and simply broken $U(2)$-symmetry~\cite{Barbieri:1995uv,Barbieri:1997tu,Roberts:2001zy,Dudas:2013pja,Falkowski:2015zwa,Linster:2018avp} determines the mixing angles between the first and the two heavier generations in terms of quark mass ratios, 
 while  giving, at the same time, a correct account of all quark masses and CKM angles in terms of two small symmetry breaking parameters $\epsilon, \epsilon^\prime$, both of order $V_{cb}$, and of ${\cal O}(1)$ factors. 
 This outcome is summarised by the forms taken by the unitary transformations that diagonalise the Yukawa couplings $Y^U$ and $Y^D$ on the left side, with a proper choice of quark phases~\cite{Dudas:2013pja,Falkowski:2015zwa,Linster:2018avp},
\begin{align}
U^L & =
\begin{pmatrix}
1&U_{12}& 0\\
- U_{12}^* &1&U_{23}\\
U_{12}^*U_{23}^* &- U_{23}^* &1
\end{pmatrix} \, , &
D^L & =
\begin{pmatrix}
1&D_{12}& D_{13}\\
- D_{12}^* &1&D_{23}\\
D_{12}^*D_{23}^* - D_{13}^*&- D_{23}^* &1
\end{pmatrix} \, , 
\label{UDL}
\end{align}
where
\begin{align}
|U_{12}| & =\sqrt{\frac{m_u}{m_c}} \, , &
|D_{12}| & =\sqrt{\frac{m_d}{m_s}}\sqrt{c_d} \, , &
|D_{13}| & =\sqrt{\frac{m_dm_s}{m_b^2}}\frac{s_d}{\sqrt{c_d}} \, , 
\label{quark_ratios}
\end{align}
and 
\begin{align}
U_{23},D_{23} & = {\cal O} (\epsilon) \, , &
\tan(\theta_d)  & \equiv |Y^{D}_{32}/Y^{D}_{33}| \, , 
& c_d & = \cos(\theta_d) \, , & s_d & = \sin(\theta_d) \, .
\end{align}
These relations are valid up to relative corrections of order $m_u/m_c$ in the up-sector and of order $m_d/m_s$ in the down sector.

Similarly, with an extended analogous definition of $U(2)$ on the leptons, the matrix $E^L$ that diagonalises the charged lepton Yukawa coupling $Y^E$ on the left side has the same form of $D^L$ with
\begin{align}
|E_{12}| & =\sqrt{\frac{m_e}{m_\mu}}\sqrt{c_e} \, , & 
|E_{13}| & =\sqrt{\frac{m_em_\mu}{m_\tau^2}}\frac{s_e}{\sqrt{c_e}} \, , & 
\tan(\theta_e) &   \equiv |Y^{E}_{32}/Y^{E}_{33}| \, .
\label{lepton_ratios}
\end{align}
and $E_{23} = {\cal O} (\epsilon)$.

Let us now turn to  B-decays, with possible anomalies  due to the exchange of a  vector leptoquark $V_\mu^a$, transforming as 
\begin{equation}
V_\mu^a= (\mathbf{3},1)_{2/3}
\end{equation}
under the SM gauge group. To make these anomalies observable in current or foreseen experiments, 
$V_\mu^a$ cannot be coupled universally to the three generations of quarks and leptons, since its exchange would lead to a branching ratio for $K_L\rightarrow \mu e$ far bigger than the current bound. To address this problem we assume that $V_\mu^a$  is coupled universally to three generations of heavy Dirac fermions, $F= Q,L,U,D,E$, with the same quantum numbers of the usual multiplets $f = q, l, u, d, e$ under the SM gauge group, mixed with $f$ by gauge invariant  bilinear mass terms. A key point is the distinction between the $F$'s and the $f$'s. This can be either because the $F$'s are composite, like $V_\mu^a$ itself, whereas the $f$'s are elementary~\cite{Barbieri:2016las,Diaz:2017lit}, or because the $F$'s  transform non-trivially under an extra gauge group, which does not act on the light fermions $f$~\cite{DiLuzio:2017vat}.

The question that we ask in this work  is whether the flavour symmetry responsible for the above relations can be extended to $V_\mu^a$ and $F$ in such a way that the violation of LFU in $B$-decays is controlled  by a minimum number of parameters - in fact the same $\epsilon, \epsilon^\prime$ and ${\cal O} (1)$ coefficients referred to above - without (or with a minimum of) {\it ad hoc} hypotheses\footnote{For a recently proposed alternative, also compatible with a suitable $U(2)$-symmetry, see Ref.~\cite{Gherardi:2019zil}.}. In view of the still  evolving character of the data on LFU in $B$-decays, we ask this question without explicitly aiming at reproducing the current values of the putative anomalies. We think that the  precision foreseen in future measurements~\cite{Bediaga:2018lhg,Kou:2018nap,Cerri:2018ypt,Tonelli} justifies this attitude.

\section{Leptoquark interactions}
\label{def}

Referring to Section~\ref{existenceproof} for an explicit realization, here we assume that the bridging alluded to in the last paragraph of the Introduction is possible, so as to see its general consequences. 
In synthetic notation the reference Lagrangian, invariant under the   SM gauge group, is
\begin{equation}
\mathcal{L} = \mathcal{L}_{\rm kin} +M_V^2 V_\mu V_\mu^\dagger + (\bar{F}M_F F +m \bar{F}\lambda_{\rm mix} f +
v \bar{f}^c \lambda_Y f + {\rm h.c.}) + \mathcal{L}_{\rm int} \, , 
\end{equation}
where $\mathcal{L}_{\rm kin}$ includes the gauge invariant interactions of $f, F$ and $V_\mu^a$ with the SM gauge bosons, and $\mathcal{L}_{\rm int}$ has the  form

\begin{equation}
\mathcal{L}_{\rm int}= g_V V_\mu^a ( \bar{Q}^a_i\gamma_\mu L_i + \bar{D}^a_i\gamma_\mu E_i)+ {\rm h.c.}
\label{Lint}
\end{equation}
where $i=1,2,3$ is the flavour index, left implicit in the fermion mass bilinear terms. Note that the leptoquark does only interact with the heavy fermions $F$ but not  with the light fermions $f$ because of their different nature, as emphasised above.
The matrices $M_F, \lambda_{\rm mix}$ and $\lambda_Y$ act in gauge and flavour space. We take all the usual multiplets in $f$ as left-handed, so that the heavy $F$ in the mixing term are only the right-handed components. We do not include right-handed neutrinos, assumed to be heavy. In the heavy sector we assume flavour universality of the mass matrix $M_F$ and of the  leptoquark interactions in  $\mathcal{L}_{\rm int}$. The flavour independence of $M_F$ is a purely simplifying assumption that does not affect any of our equations, whereas the universality of $\mathcal{L}_{\rm int}$ helps in reducing the number of free parameters. This assumption, however, is well justified in concrete examples, either in strongly interacting composite Higgs models, where flavour could be associated with an approximate global symmetry, like in QCD, or if $\mathcal{L}_{\rm int}$ arises from an extended gauge interaction of the heavy $F$'s, which is universal by construction.

To determine the leptoquark interactions with the light fermion eigenstates, it is useful to first go to the diagonal basis of 
$m \bar{F}\lambda_{\rm mix} f $ by proper unitary transformations of the $F$ and the $f$ fields. In general the transformations of the heavy fields, being different  for $Q$ and $L$, as well as for $D$ and $E$, introduce  unitary matrices in $\mathcal{L}_{\rm int}$, eq.~(\ref{Lint})~\cite{DiLuzio:2018zxy}. Keeping the same notation for the rotated fields, in the new basis the interaction Lagrangian becomes
\begin{equation}
\mathcal{L}_{\rm int}\rightarrow g_V V_\mu^a ( \bar{Q}^a\gamma_\mu V^{LQ} L + \bar{D}^a\gamma_\mu V^{DE}E) + {\rm h.c.}
\label{Lint'}
\end{equation}
Given the  diagonal form of the mixing matrices   $m_{q,l}$ and $m_{d,e}$ in the new basis, it is easy to extract the light 
fermions, massless in the limit of unbroken electroweak symmetry, in the normalised combinations
\begin{align}
q^\prime & =\hat{c}_q q- \hat{s}_q Q_L \, , &
l^\prime & =\hat{c}_l l-\hat{s}_l L_L \, ,  &
 d^\prime & =\hat{c}_d d- \hat{s}_d D_L \, , &
e^\prime & =\hat{c}_e e- \hat{s}_e E_L \, ,
\end{align}
where $\hat{s}_{q} (\hat{c}_{q})$ are sines (cosines) of mixing angles with the same diagonal form and typical size of  order $m_{q}/M_Q$, and similarly for the other angles.

For the purposes of the present section, to be justified later on in Section \ref{existenceproof}, we assume that the (broken) flavour symmetry implies  for all the elements of $\hat{s}_{d,e}$ that they be sufficiently small, 
\begin{equation}
(\hat{s}_{d,e})_{ii} \lesssim {\cal O}(\epsilon^2) \, .
\label{s_deii}
\end{equation}
As it can be explicitly checked quantitatively for all the appropriate observables, this implies that the only phenomenologically relevant   interaction of the leptoquark with the light fields, 
omitting the primed indices, 
\begin{equation}
\mathcal{L}_{\rm int}^{\rm light \, fields}= g_V V_\mu^a ( \bar{q}^a\gamma_\mu \hat{s}_q V_{QL}\hat{s}_l l) + {\rm h.c.}
\label{Lint_light}
\end{equation}
Finally, in terms of the  unitary transformations $U^L, D^L, E^L$ that diagonalise on the left side the Yukawa couplings of the up- and down-quarks and the charged leptons respectively\footnote{
The diagonalisation of the mixing terms leads to a modification of the Yukawa couplings $\lambda_Y\rightarrow \hat{\lambda}_Y$. One can show that $\hat{\lambda}_Y$ differs from $\lambda_Y$ by ${\cal O}(1)$ factors and by sub-leading corrections in $\epsilon,\epsilon^\prime$, thus not affecting the forms of eqs.~\eqref{UDL}, \eqref{quark_ratios}, \eqref{lepton_ratios}.}, the final expression for the interaction Lagrangian in the physical mass basis is
\begin{equation}
\mathcal{L}_{\rm int}^{\rm physical}= g_V  V_\mu^a ( \bar{d}_{L}^a\gamma_\mu F^De_{L}+ \bar{u}_{L}^a\gamma_\mu F^U\nu_{L}) + {\rm h.c.}
\label{Lint_light_final}
\end{equation}
where 
\begin{equation}
F^D= D^{ L \dagger}\hat{s}_q V_{QL}\hat{s}_l E^L \, , \quad\quad F^U= U^{ L \dagger}\hat{s}_q V_{QL}\hat{s}_l E^L \, .
\end{equation} 
Note that the transformation $V_{QL}\rightarrow e^{i\Phi_Q}V_{QL}e^{i\Phi_L}$, with $e^{i\Phi_{Q,L}}$ diagonal phase matrices, can be reabsorbed by proper phase redefinitions of $U^L, D^L, E^L$ and of the light fields without changing the form of eqs.~\eqref{UDL}, \eqref{quark_ratios}, \eqref{lepton_ratios} nor the CKM matrix $V_{\rm CKM}=U^{ L \dagger}D^L$.
Using this phase freedom, if we further require from the flavour symmetry, to be justified later on in Section \ref{existenceproof}, that 
\begin{equation}
(\hat{s}_{q,l})_{11} \lesssim {\cal O}(\epsilon^2), \quad\quad (\hat{s}_{q,l})_{22} \equiv s_{q2,l2}\lesssim {\cal O}(\epsilon),
\label{sql11}
\end{equation}
$V_{QL}$ can be effectively reduced, in the cases to be considered below,  to a real rotation between the second and the third generation, defined by an angle $\theta_{ql}$ ($c_{ql}=\cos\theta_{ql}, s_{ql}=\sin\theta_{ql}$).

\section{Violations of Lepton Flavour Universality}
\subsection{General expressions}

By integrating out the leptoquark from (\ref{Lint_light_final}), one obtains the effective Lagrangians relevant to describe the LFU violations:
\begin{align}
\mathcal{L}_{\rm eff}^{\rm CC} & = -\left(\frac{g_V}{M_V} \right)^2 F^{D*}_{b\tau} F^U_{c\tau} (\bar{c}_{L}\gamma_\mu b_L)
(\bar{\tau}_{L}\gamma_\mu  \nu_{3L}) \, , \\
\mathcal{L}_{\rm eff}^{\rm NC} & = -\left(\frac{g_V}{M_V} \right)^2 F^{D*}_{b\mu}  F^D_{s\mu} (\bar{s}_{L}\gamma_\mu b_L)
(\bar{\mu}_{L}\gamma_\mu  \mu_{L}) \, .
\end{align}
Therefore, 
from the usual definition,
\begin{equation}
R_{D^{(*)}}\equiv \frac{BR(B\rightarrow D^{(*)}\tau\nu)}{BR(B\rightarrow D^{(*)} l \nu)},~~ l=e, \mu,
\end{equation}
one has
\begin{equation}
\Delta R_{D}\equiv \frac{R_{D^{(*)}}}{R_{D^{(*)}}^{SM}}-1= \left(\frac{g_V}{M_V}\right)^2\frac{1}{\sqrt{2}G_F}\mathcal{R}e \left(\frac{F^{D*}_{b\tau} F^U_{c\tau}}{V_{cb}}\right)\, , 
\end{equation}
where we neglect suppressed contributions that do not interfere with the SM amplitude.

Similarly, encapsulating the neutral current anomaly into the Wilson coefficient $\Delta C_9^\mu$ as usually done in the literature $(\Delta C^\mu_{10}= -\Delta C^\mu_{9})$\footnote{For the theoretically clean observables $\Delta R_K\equiv 1-
R_K|_{[1,6] \GeV^2}$ and $\Delta R_{K^*}\equiv 1-
R_{K^*}|_{[1.1,6] \GeV^2}$, it is $\Delta R_{K} \approx \Delta R_{K^*}\approx -0.46 \Delta C^\mu_9$~\cite{Capdevila:2017bsm}.}
\begin{equation}
\mathcal{L}_{\rm eff}^{\rm NC} = 4\sqrt{2}G_FV_{tb}V_{ts}^*\frac{\alpha}{4\pi}\Delta C_9^\mu 
(\bar{s}_{L}\gamma_\mu b_L)
(\bar{\mu}_{L}\gamma_\mu \mu_L) \, , 
\end{equation}
one has
\begin{equation}
\Delta C_9^\mu = - \left(\frac{g_V}{M_V}\right)^2\frac{4\pi}{\alpha}\frac{1}{4\sqrt{2} G_F}
\mathcal{R}e\left(\frac{F^{D*}_{b\mu} F^D_{s\mu}}{V_{tb}V_{ts}^*}\right) \, .
\end{equation}
These expressions do not depend on the phases of the fermion fields, as they have to.
Using the expressions for $U^L, D^L, E^L$ in Section \ref{intro} with their phase convention and expanding in $\epsilon$, it is
\begin{align}
F^D_{b\tau}&\approx s_{q3}s_{l3}c_{ql} \simeq {\cal O}(1) \, , \label{a1} \\
F^U_{c\tau}&\approx s_{q3}s_{l3}\left(-c_{ql}U_{23} + s_{ql}\frac{s_{q2}}{s_{q3}} \right) \simeq {\cal O}(\epsilon)\, , 
\label{a2} \\
F^D_{b\mu}&\approx s_{q3}s_{l3} \left(-c_{ql}E_{23}^* - s_{ql}\frac{s_{l2}}{s_{l3}}\right)\simeq {\cal O}(\epsilon)\, , 
\label{a3} \\
F^D_{s\mu}&\approx s_{q3}s_{l3}\left(c_{ql}D_{23} E_{23}^* +c_{ql}\frac{s_{q2}}{s_{q3}}\frac{s_{l2}}{s_{l3}}
- s_{ql}\frac{s_{q2}}{s_{q3}}E_{23}^*+s_{ql}\frac{s_{l2}}{s_{l3}}D_{23} \right) \simeq {\cal O}(\epsilon^2) \, .
\label{a4}
\end{align}
At the same time one has
\begin{equation}
V_{cb}\approx - V_{ts}^*\approx D_{23}-U_{23} \simeq {\cal O}(\epsilon) \, .
\label{Vcb}
\end{equation}. 

\subsection{Expected range for  LFU violations}
\subsubsection{ Minimal model}
\label{epsilon2}

\begin{figure}[t]
\centering
\includegraphics[clip,width=.55\textwidth]{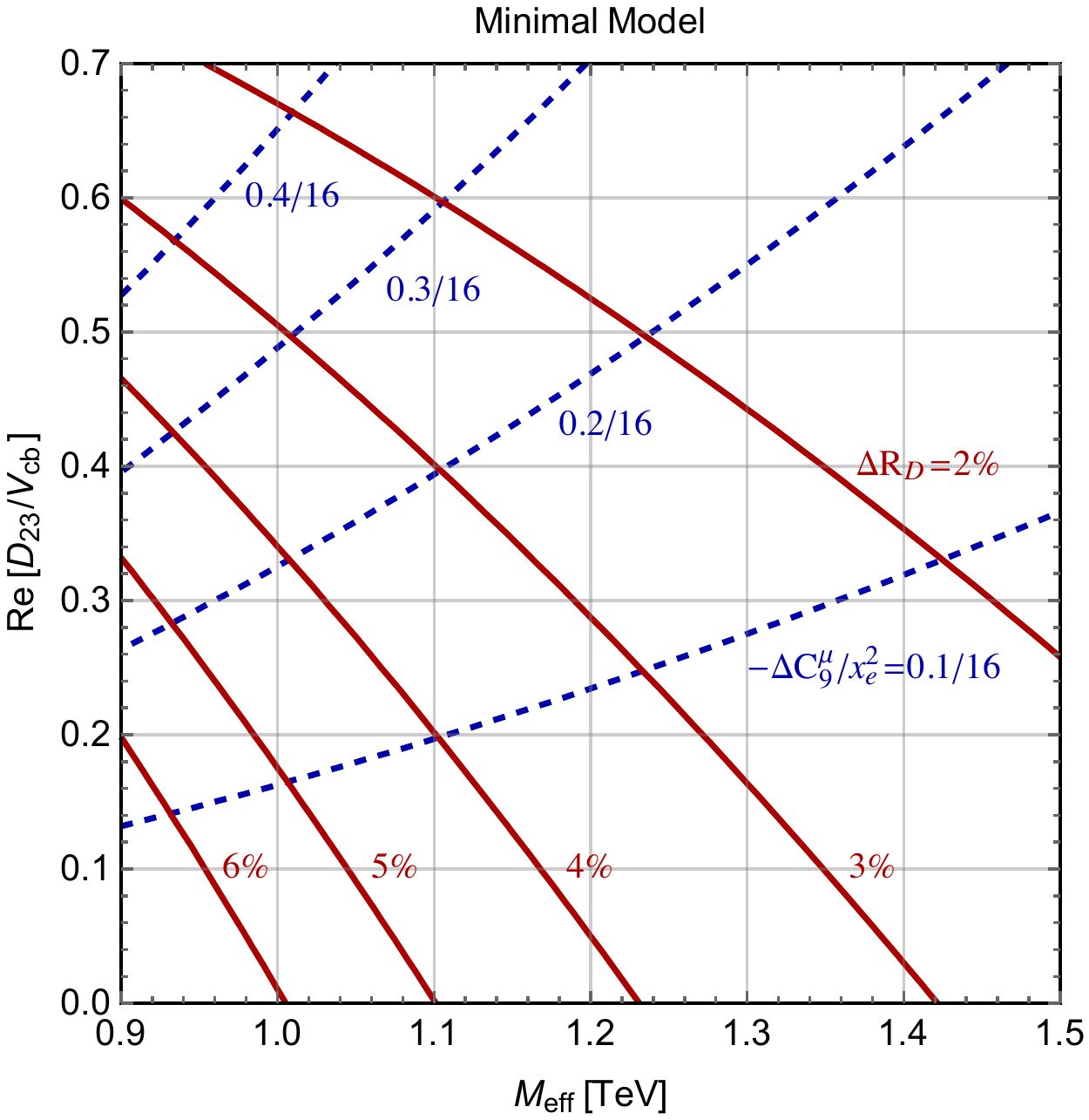}
\caption{Isolines of the charged current (CC, red solid lines) and of the neutral current (NC, blue dashed lines) anomaly in the minimal model for
$\Delta R_{D}= \{2, 3, 4, 5, 6\}\%$ and $-\Delta C^\mu_9/x_e^2= \{0.1/16, 0.2/16, 0.3/16,0.4/16\}$ respectively, and $x_e=|E_{23}/V_{cb}|$.}
\label{fig:D32<U32}
\end{figure}

A strong simplification occurs in eqs.~(\ref{a1}-\ref{a4}) if $ s_{q2,l2} \lesssim {\cal O}(\epsilon^2)$, to be justified in Section \ref{MM}, so that each $F^{U,D}_{ij}$ is dominated by the first terms on the r.h.s. of these equations. In this case
\begin{equation}
\mathcal{R}e\left(\frac{F^{D*}_{b\tau} F^U_{c\tau}}{V_{cb}}\right)=
-\left(s_{q3}s_{l3}c_{ql}\right)^2  \mathcal{R}e\left(\frac{U_{23}}{V_{cb}}\right) \, ,
\end{equation}
and
\begin{equation}
\mathcal{R}e \left(\frac{F^{D*}_{b\mu} F^D_{s\mu}}{V_{tb}V_{ts}^*} \right)=
\left(s_{q3}s_{l3}c_{ql} \right)^2 |E_{23}|^2 \mathcal{R}e\left(\frac{D_{23}}{V_{cb}}\right) \, , 
\end{equation}
so that, from eq.~(\ref{Vcb}),
\begin{align}
\Delta R_{D} & =\left( \frac{g_Vs_{q3}s_{l3}c_{ql}}{M_V}\right)^2\frac{1}{\sqrt{2}G_F}
\left[ 1-\mathcal{R}e \left(\frac{D_{23}}{V_{cb}} \right)\right] \nonumber \\
& = 0.06 \left( \frac{\TeV}{M_{\rm eff}} \right)^2  \left[ 1-\mathcal{R}e \left(\frac{D_{23}}{V_{cb}} \right)\right]\, , 
\end{align}
with $M_{\rm eff} \equiv M_V/(g_V s_{q3} s_{l3} c_{ql})$ and
\begin{align}
\Delta C_9^\mu & = - \left( \frac{g_Vs_{q3}s_{l3}c_{ql}}{M_V}\right)^2\frac{4\pi}{\alpha}\frac{1}{4\sqrt{2} G_F}
|E_{23}|^2 \mathcal{R}e\left( \frac{D_{23}}{V_{cb}} \right) \nonumber \\
& = -0.04 \left( \frac{\TeV}{M_{\rm eff}} \right)^2  \left| \frac{E_{23}}{V_{cb}} \right|^2 \mathcal{R}e\left( \frac{D_{23}}{V_{cb}} \right) \, .
\end{align}
\begin{figure}[t]
\centering
\includegraphics[clip,width=.55\textwidth]{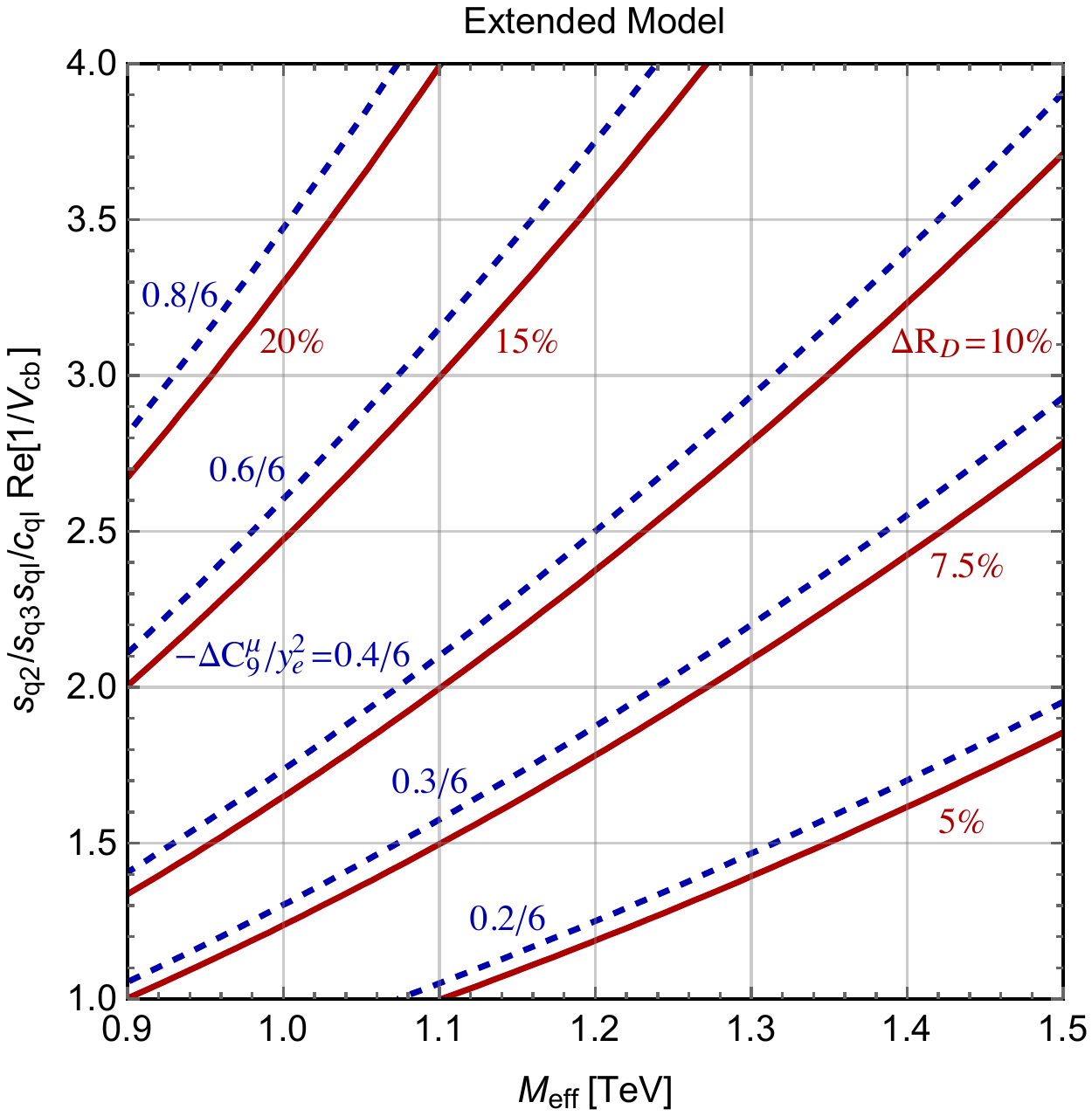}
\caption{Isolines of the charged current (CC, red solid lines) and of the neutral current (NC, blue dashed lines) anomaly in the extended model for
$\Delta R_{D}= \{ 5, 7.5, 10, 15, 20\}\%$ and $-\Delta C^\mu_9/y_e^2=\{ 0.2/6, 0.3/6, 0.4/6, 0.6/6, 0.8/6 \}$ respectively, with $y_e |V_{cb}|=s_{l2}/s_{l3}$.}
\label{fig:anomalies}
\end{figure}
The two anomalies are represented in Fig.~\ref{fig:D32<U32} in a range of values for $M_{\rm eff}$  compatible with current bounds from direct searches of the leptoquark in  pair production, $pp\rightarrow V V^\dagger$, and indirect searches via $pp\rightarrow \tau \bar{\tau}$~\cite{Sirunyan:2018vhk,Aaboud:2017sjh,Sirunyan:2018lbg,Faroughy:2016osc, Angelescu:2018tyl,Schmaltz:2018nls,Baker:2019sli}. Especially in the CC case, the values of the anomalies in Fig.~\ref{fig:D32<U32}  are definitely lower than the central values of the current averages~\cite{Altmannshofer:2017yso,Capdevila:2017bsm,Caria,Alguero:2019ptt,Aebischer:2019mlg,Blanke:2019qrx,Shi:2019gxi}
\begin{equation}
\Delta R_{D}= \left( 14 \pm 4 \right)\% \, , \qquad \qquad
\Delta C_9^\mu = - \left( 0.53 \pm 0.09 \right) \, ,
\label{CCdata}
\end{equation}
which are, however, still evolving and have relatively large errors. These values, however, are not outside the expected sensitivity of future experiments~\cite{Bediaga:2018lhg,Kou:2018nap,Cerri:2018ypt}, eventually with a modest improvement in the theory.

\subsubsection{ Extended model}
\label{epsilon}

More parameters are involved if $s_{q2,l2}={\cal O}(\epsilon)$.  We consider $s_{lq}={\cal O}(1)$ and, in order to represent this case, although with a corresponding uncertainty, among the ${\cal O}(\epsilon)$ parameters we take $s_{q2}/s_{q3}$ and  $s_{l2}/s_{l3}$ dominant over $|U_{23}|, 
|D_{23}|, |E_{23}|$. This gives
\begin{equation}
\mathcal{R}e \left(\frac{F^{D*}_{b\tau} F^U_{c\tau}}{V_{cb}}\right)\approx
\left(s_{q3}s_{l3}c_{ql} \right)^2  \frac{s_{q2}}{s_{q3}}  \frac{s_{ql}}{c_{ql}}  \mathcal{R}e \left(\frac{1}{V_{cb}} \right) \, ,
\end{equation}
and
\begin{equation}
\mathcal{R}e\left(\frac{F^{D*}_{b\mu} F^D_{s\mu}}{V_{tb}V_{ts}^*}\right)\approx
\left(s_{q3}s_{l3}c_{ql}\right)^2   \frac{s_{ql}}{c_{ql}}\frac{s_{q2}}{s_{q3}}  \left(\frac{s_{l2}}{s_{l3}} \right)^2 \mathcal{R}e \left(\frac{1}{V_{cb}} \right) \, , 
\end{equation}
so that
\begin{align}
\Delta R_D & = 0.06 \left( \frac{\TeV}{M_{\rm eff}} \right)^2 \left( \frac{s_{q2} s_{ql}}{s_{q3} c_{ql}} \frac{1}{\mathcal{R}e (V_{cb})} \right) \, , \\
\Delta C^\mu_9 & = - 0.04 \left( \frac{\TeV}{M_{\rm eff}} \right)^2 \left( \frac{s_{q2} s_{ql}}{s_{q3} c_{ql}} \frac{1}{\mathcal{R}e (V_{cb})} \right) \left(\frac{s_{l2}}{s_{l3}} \frac{1}{|V_{cb}|} \right)^2 \, .
\end{align}
In Fig.~\ref{fig:anomalies}  we represent the two anomalies in the range of values explicitly indicated. Unlike the previous case, these values can be close to the ones currently observed.

\section{LFU violations and flavour symmetries}
\label{existenceproof}
\subsection{Relating mixing angles to fermion masses}
\label{predictions}

As anticipated in the Introduction, for the ease of the reader we recall 
the two  ingredients needed to give rise to the mass-angle relations in eqs.~\eqref{UDL}, \eqref{quark_ratios}, \eqref{lepton_ratios}:
\begin{itemize}
\item An $SU(2)_f\times U(1)_f$ symmetry that acts as $U(2)$ on the first two generations, one doublet for any irreducible  representation of  the SM gauge group - $q, l, u, d, e$ in  standard notation, all left-handed Weyl spinors - and the $U(1)_f$ factor extended to act on the third generation $SU(2)$-singlets with charges given in Table \ref{U1charges}. 
 These charges, which account  for the relative heaviness of the top among the third generation particle themselves, are normalised  to the $U(1)_f$-charge of the first two generation doublets, transforming as $\bold{2}_1$ under $SU(2)_f\times U(1)_f$.
\item Two {\it spurions}, one doublet and one singlet under $SU(2)_f\times U(1)_f$
\begin{align}
\Sigma & = \bold{2}_{-1}= 
\begin{pmatrix}
\epsilon \Lambda_f \\
0
\end{pmatrix} \, , 
& \chi & = \bold{1}_{-1} = \epsilon^\prime \Lambda_f \, , 
\label{spurions}
\end{align}
where $\Lambda_f$ is the UV scale of the flavour sector, i.e. the scale at which the spurions enter as scalar fields into an effective
$SU(2)_f\times U(1)_f$-invariant Lagrangian, and, without loss of generality,  we have taken $\Sigma$  pointing in the first direction. The dimensionless parameter $\epsilon$ is of order of $V_{cb}$ and $\epsilon^\prime$ is a  factor of a few times smaller than $\epsilon$. Their determination is not precise, since it depends on the unknown ${\cal O}(1)$ factors that are allowed to enter  the effective Lagrangian.
\end{itemize}
 \begin{table}[t]\small
$$\begin{array}{c|c|c|c|c|c}
&q_3&u_3&d_3&l_3&e_3\\ \hline
U(1)_f&0&0&1&0&1
\end{array}$$
\caption{\label{U1charges}
$U(1)_f$ charges of the third generation fermions, which are $SU(2)_f$ singlets. The first two generations all transform as $\bold{2}_1$ under $SU(2)_f\times U(1)_f$.
}
\label{U1charges}
\end{table}
Eqs.~\eqref{UDL}, \eqref{quark_ratios}, \eqref{lepton_ratios} arise from the most general Yukawa couplings $Y^{U,D,E}(\Sigma, \chi; \Lambda_f)$ consistent with the $SU(2)_f\times U(1)_f$ symmetry and ${\cal O}(1)$ parameters\footnote{Refs.~\cite{Dudas:2013pja,Falkowski:2015zwa,Linster:2018avp} consider the case with the $U(1)_f$-charges of $l_3$ and $e_3$ interchanged with respect to the ones in Table~\ref{U1charges}, thus commuting with the $SU(5)$ generators. While this choice leaves eqs. (\ref{UDL},\ref{quark_ratios}) unchanged, it would suppress to  $O(\epsilon$) the leptoquark interactions to the third generation fermions.}.
From $V_{\rm CKM}=U^{ L \dagger}D^L$ and suitable choices of the quark phases, eqs.~\eqref{UDL} and \eqref{quark_ratios}  lead to the relations
\begin{align}
V_{us}&=\left| \sqrt{\frac{m_d}{m_s}}\sqrt{c_d}-
e^{-i\alpha_1}\sqrt{\frac{m_u}{m_c}} \right|
\label{Vus} \, , \\
V_{td}&=e^{i\tilde{\alpha}_1}\sqrt{\frac{m_d}{m_s}}\sqrt{c_d}\left(|V_{cb}| -
e^{i\alpha_2}\frac{s_d}{c_d}\frac{m_s}{m_b} \right)
\label{Vtd} \, , \\
V_{ub}&=-e^{-i(\alpha_1+\tilde{\alpha}_1)} \left(\sqrt{\frac{m_u}{m_c}}|V_{cb}| -
e^{i(\alpha_1-\alpha_2)}\sqrt{\frac{m_d}{m_s}}\frac{s_d}{\sqrt{c_d}}\frac{m_s}{m_b} \right) \, , 
\label{Vub}
\end{align}
where
\begin{equation}
\tilde{\alpha}_1= \arg \left[\sqrt{\frac{m_d}{m_s}}\sqrt{c_d}-
e^{-i\alpha_1}\sqrt{\frac{m_u}{m_c}} \, \right] \, .
\end{equation}
Table~\ref{SU2-predictions} shows the predictions of $U(2)$ models with $\theta_d = 0$~\cite{Barbieri:1995uv,Barbieri:1997tu} compared with the current experimental values, using the CKM input from Ref.~\cite{UTfit}. 
Clearly these data, in particular the value of $V_{ub}/V_{cb}$,  favor $U(2)$ models with $\theta_d\neq 0$~\cite{Roberts:2001zy, Dudas:2013pja,Falkowski:2015zwa,Linster:2018avp}. Indeed all relations above are brought to precise agreement with data, including the CP violating phase, for either $c_d= 0.91\pm 0.03$, $\alpha_1=-1.6\pm 0.2$, $\alpha_2=1.5\pm 0.1$, or $c_d= 0.66\pm 0.04$, $\alpha_1=2.6\pm 0.3$, $\alpha_2=1.5\pm 0.1$.
 \begin{table}[h]
$$\begin{array}{c|c|c}
|V_{us}|&|V_{td}/V_{cb}|&|V_{ub}/V_{cb}|      \\ \hline
0.16 \div 0.29&0.22 (2)&0.045 (9) \\ \hline
0.2251(6)&0.21(1)&0.093(6)
\end{array}$$
\caption{$U(2)$ predictions for $s_d=0$ (second line) and current experimental values (third line). With $\theta_d\neq 0$ all these relations,  in particular the one for $V_{ub}/V_{cb}$,  are brought to precise agreement with data.}
\label{SU2-predictions}
\end{table}

Can one extend this flavour symmetry to  the heavy fermions $F$ in a way consistent with eq.~(\ref{Lint}) and such that the conditions (\ref{s_deii}) and (\ref{sql11}) are automatically satisfied? We show that the answer is positive, distinguishing the two cases considered in Sections~\ref{epsilon2} and \ref{epsilon}, respectively called Minimal Model and Extended Model.

\subsection{ Minimal model}
\label{MM}
 Under $SU(2)_f\times U(1)_f$ we assume that the heavy Dirac fermions $F= Q,L,U,D,E$ transform  as the charge conjugated of the corresponding $f = q, l, u, d, e$ with the $U(1)_f$ charges chosen according to Table~\ref{U1charges}. Furthermore we require that the mixing terms between $F$ and $f$ respect the flavour symmetry with inclusion of the spurions $\Sigma$ and $\chi$, see eq.~(\ref{spurions}), as it is the case for the Yukawa couplings of the fermions $f$ themselves.

In full generality the mixing mass terms acquire the form:
\begin{equation}
\begin{split}
\mathcal{L}_{\rm mixing}(Q,q) &=m[\bar{Q}_3q_3 +\bar{Q}_3(\Sigma_a  \epsilon_{ab}q_b)+(\bar{Q}_a \epsilon_{ab}\Sigma_b)q_3+
(\bar{Q}_a \epsilon_{ab}\Sigma_b)(\Sigma_c  \epsilon_{cd}q_d)\\
&+(\bar{Q}_a\epsilon_{ab}q_b)\chi^2 + \bar{Q}_3(\Sigma^*_aq_a)\chi^2+(\bar{Q}_a\Sigma^*_a)q_3\chi^2 +
(\bar{Q}_a\Sigma^*_a)(\Sigma^*_b q_b)\chi^4] \, , 
\end{split}
\end{equation}
and similarly for $\mathcal{L}_{\rm mixing}(L,l)$, where in front of every term we leave understood  an ${\cal O}(1)$ factor and an appropriate inverse power of $\Lambda_f$;
\begin{equation}
\begin{split}
\mathcal{L}_{\rm mixing}(D,d) &=m[\bar{D}_3d_3\chi^2 +\bar{D}_3(\Sigma_a  \epsilon_{ab}d_b)\chi+(\bar{D}_a \epsilon_{ab}\Sigma_b)d_3\chi+
(\bar{D}_a \epsilon_{ab}\Sigma_b)(\Sigma_c  \epsilon_{cd}d_d)\\
&+(\bar{D}_a\epsilon_{ab}d_b)\chi^2 + \bar{D}_3(\Sigma^*_ad_a)\chi^3+(\bar{D}_a\Sigma^*_a)d_3\chi^3 +
(\bar{D}_a\Sigma^*_a)(\Sigma^*_b d_b)\chi^4] \, , 
\end{split}
\end{equation}
and similarly for $\mathcal{L}_{\rm mixing}(E,e)$. 

Upon use of eq.~(\ref{spurions}) one obtains  these mixing terms in matrix form:
\begin{equation}
\mathcal{L}_{\rm mixing}(Q,q)= (\bar{Q}_1,\bar{Q}_2,\bar{Q}_3) m_q 
\begin{pmatrix}
q_1 \\
q_2\\
q_3
\end{pmatrix},\quad\quad
m_q=
\begin{pmatrix}
\epsilon^2\epsilon^{\prime 4} &\epsilon^{\prime 2}&\epsilon\epsilon^{\prime 2} \\
-\epsilon^{\prime 2}&\epsilon^2&\epsilon\\
\epsilon\epsilon^{\prime 2}&\epsilon&1
\end{pmatrix},
\end{equation}
(again with ${\cal O}(1)$ factors left understood) and similarly for $\mathcal{L}_{\rm mixing}(L,l)$ with a matrix $m_l$. In the same way
\begin{equation}
\mathcal{L}_{\rm mixing}(D,d)= (\bar{D}_1,\bar{D}_2,\bar{D}_3) m_d 
\begin{pmatrix}
d_1 \\
d_2\\
d_3
\end{pmatrix},\quad\quad
m_d=
\begin{pmatrix}
\epsilon^2\epsilon^{\prime 4} &\epsilon^{\prime 2}&\epsilon\epsilon^{\prime 3} \\
-\epsilon^{\prime 2}&\epsilon^2&\epsilon\epsilon^{\prime}\\
\epsilon\epsilon^{\prime 3}&\epsilon\epsilon^{\prime}&\epsilon^{\prime 2}
\end{pmatrix},
\end{equation}
as for $\mathcal{L}_{\rm mixing}(E,e)$ with a matrix $m_e$.
Note that, by gauge invariance, the heavy fermions in $\mathcal{L}_{\rm mixing}$ are all only right-handed whereas in $\mathcal{L}_{\rm int}$, eq.~(\ref{Lint}), they are fully Dirac fields.

Following Section~\ref{def}, of particular relevance are the diagonal forms of $m_{q,l}$ and $m_{d,e}$ in the new bases
\begin{equation}
m_{q,l} = m 
\begin{pmatrix}
\epsilon^{\prime 4}/\epsilon^2 &0&0 \\
0&\epsilon^2&0\\
0&0&1
\end{pmatrix},\quad\quad
m_{d,e} = m 
\begin{pmatrix}
\epsilon^{\prime 4}/\epsilon^2 &0&0 \\
0&\epsilon^2&0\\
0&0&\epsilon^{\prime 2} \, , 
\end{pmatrix}
\end{equation}
with ${\cal O}(1)$ factors, different for $q,l,d,e$, left understood. As desired, this automatically implies eqs.~(\ref{s_deii}) and (\ref{sql11}) with, in particular, $s_{q2,l2} \simeq {\cal O}(\epsilon^2)$. The form of $m_{q,l}$ also shows that, in this case, $s_{ql}={\cal O}(\epsilon)$.

\subsection{Extended model: an existence proof}

 \begin{table}[t]\small
$$\begin{array}{ccccccc|c}
&\mathcal{Q}_a&\mathcal{Q}_3&\mathcal{U}_a&\mathcal{U}_3&\mathcal{D}_a&\mathcal{D}_3&\Sigma^F\\ \hline
SU(2)_F&\bar{\bold{2}}&\bold{1}&\bar{\bold{2}}&\bold{1}&\bar{\bold{2}}&\bold{1}&\bold{2}\\ \hline
U(1)_F&-1&0&0&-1&0&-1&-1
\end{array}$$
\caption{Transformation properties under $SU(2)_F\times U(1)_F$ of the heavy fermions, grouped in $SU(4)$ multiplets
as in eq.~(\ref{multiplets}).}
\label{SU2-U1charges}
\end{table}

To reproduce the conditions of Fig.~\ref{fig:anomalies}, we need $ s_{q2,l2} = {\cal O}(\epsilon)$ as well as $s_{ql}={\cal O}(1)$.
To define the flavour symmetry, let us first organise the heavy fermions $F$ into quartets of $SU(4)$, as it has been the case for the light fermions $f$:
\begin{equation}
\mathcal{Q}_i=
\begin{pmatrix}
Q \\
L
\end{pmatrix}_i \, ,\quad\quad
\mathcal{U}_i=
\begin{pmatrix}
U\\
N
\end{pmatrix}_i \, ,\quad\quad
\mathcal{D}_i=
\begin{pmatrix}
D\\
E
\end{pmatrix}_i\, ,
\label{multiplets}
\end{equation}
with $i=1,2,3$ a flavour index\footnote{$N$ is a Dirac fermion singlet which does not play any role in the following since we rely on the usual see-saw mechanism for neutrino masses and the mixing of $N$ with the "elementary" super-heavy Majorana $\nu_R$ leaves no light state in the $(N_L, N_R,\nu_R)$ sector.}.  We then introduce a new $SU(2)_F\times U(1)_F$ which acts on these multiplets, each split into doublets, $i\equiv a=1,2$, and singlets, $i=3$, under $SU(2)_F$.
The $U(1)_F$-charges are indicated in Table~\ref{SU2-U1charges}, where we have also included a {\it spurion} $\Sigma^F$.  We take $\Sigma^F$  pointing  in the first direction, without loss of generality, and with a vev of order $\Lambda_F$.

This choice of the $U(1)_F$ charges, admittedly {\it ad hoc} but possible, introduces mixing only in the $(Q, q)$ and $(L, l)$ sectors. Leaving ${\cal O}(1)$ factors and inverse powers of $\Lambda_F$ understood,  the most general mixing mass term in this case is
\begin{equation}
\begin{split}
\mathcal{L}_{\rm mixing}(Q,q) &=m[\bar{Q}_3q_3 +\bar{Q}_3(\Sigma_a  \epsilon_{ab}q_b)+ \bar{Q}_3(\Sigma^*_aq_a)\chi^2\\
&+ (\bar{Q}_a \epsilon_{ab}\Sigma^F_b)q_3+
(\bar{Q}_a \epsilon_{ab}\Sigma^F_b)(\Sigma_c  \epsilon_{cd }q_d)
+(\bar{Q}_a \epsilon_{ab}\Sigma^F_b)(\Sigma^*_c q_c)\chi^2] \, , 
\end{split}
\end{equation}
and similarly for $\mathcal{L}_{\rm mixing}(L,l)$. In matrix notation, with 
\begin{equation}
\Sigma^F = 
\begin{pmatrix}
\epsilon^F \Lambda_f \\
0
\end{pmatrix} \, ,
\end{equation}
it is
\begin{equation}
\mathcal{L}_{\rm mixing}(Q,q)= (\bar{Q}_1,\bar{Q}_2,\bar{Q}_3) 
\begin{pmatrix}
0 &0&0 \\
\epsilon^F \epsilon \epsilon^{\prime}&\epsilon^F\epsilon&\epsilon^F\\
\epsilon\epsilon^{\prime 2}&\epsilon&1
\end{pmatrix}
\begin{pmatrix}
q_1 \\
q_2\\
q_3
\end{pmatrix} \, .
\end{equation}
After diagonalisation, for $\epsilon^F = {\cal O}(1)$, one gets $s_{q3} \simeq {\cal O}(1), s_{q2} \simeq {\cal O}(\epsilon), s_{q1}=0$ and similarly for 
$s_{li}$,  and moreover $s_{ql} \simeq {\cal O}(1)$.

\section{Other flavour observables}
\label{other_obs}

Both in the Minimal  and in the Extended Model,  a relatively precise description of  the leptoquark couplings to the first two generations allows to predict a number of flavour-violating observables. We briefly discuss some of them in the following, with results summarized in Tables~\ref{Kmue} and \ref{DeltaB=2}.

\subsection{  $K_L\rightarrow \mu e $}

The effective Lagrangian relevant to $K_L\rightarrow \mu^- e^+$ is
\begin{equation}
\mathcal{L}=(C_{d\bar{s}} \bar{s}_L\gamma_\mu d_L + C_{s\bar{d}} \bar{d}_L\gamma_\mu s_L)
(\bar{\mu}_L\gamma_\mu e_L) +{\rm h.c.}  
\end{equation}
from which the corresponding decay amplitude is (neglecting small CP violating effects)
\begin{equation}
A(K_L\rightarrow \mu^- e^+) = C_{K\rightarrow \mu e}
<\mu^- e^+| ( \bar{s}_L\gamma_\mu d_L)(\bar{\mu}_L\gamma_\mu e_L)|\bar{K}_0> \, , 
\end{equation}
where 
\begin{equation}
C_{K\rightarrow \mu e}= \frac{1}{\sqrt{2}}(C_{d\bar{s}}e^{-i\beta}+ C_{s\bar{d}}e^{i\beta}) \, , 
\end{equation}
and $\beta$ is the phase of $V_{us}^*V_{ud}$.

In the Minimal Model it is\footnote{From eq.~(\ref{UDL}) one  can see that $D^L_{31}$, and similarly  $E^L_{31}$, receive two contributions. Here we  assume for simplicity the dominance of $D_{12}D_{23} (E_{12}E_{23})$ over $D_{13} (E_{13})$, respectively. We also drop irrelevant signs in the following.}
\begin{equation}
C_{K\rightarrow \mu e}^{\rm MM}=-\frac{1}{M_{\rm eff}^2}  \left(E_{32}^{L*} E_{31}^L\right)\sqrt{2} \mathcal{R}e [D_{31}^L D_{32}^{L*} e^{-i\beta}]
\approx  \frac{1.0 \cdot 10^{-6}}{M_{\rm eff}^2} \left|\frac{D_{23}}{V_{cb}}\right|^2 \left(\frac{x_e}{4} \right)^2\sqrt{c_d c_e} \, , 
\end{equation}

In the Extended Model it is
\begin{equation}
C_{K\rightarrow \mu e}^{\rm EM}= -\frac{1}{M_{\rm eff}^2} \left(\frac{s_{q2}}{s_{q3}} \frac{s_{l2}}{s_{l3}} \right)^2 \left(E_{22}^{L*} E_{21}^L \right)\sqrt{2} \mathcal{R}e [D_{21}^L D_{22}^{L*} e^{-i\beta}]
\approx  \frac{5.1\cdot10^{-6}}{M_{\rm eff}^2}  \left(\frac{y_q}{3} \frac{y_e}{3} \right)^2 \sqrt{ c_d c_e} \, .
\end{equation}

\subsection{$\mu N\rightarrow e N$}

The effective Lagrangian relevant to $\mu-e$ conversion is
\begin{equation}
\mathcal{L}=C_{\mu-e} (\bar{d}_L\gamma_\mu \mu_L)(\bar{e}_L\gamma_\mu d_L) +{\rm h.c.}
\end{equation}
where, in the Minimal Model,
\begin{equation}
C_{\mu-e}^{\rm MM}=- \frac{1}{M_{\rm eff}^2} \left(D^{L*}_{31} E^{L}_{32})(E^{L*}_{31} D^{L}_{31} \right)
\approx \frac{1.6 \cdot 10^{-7} }{M_{\rm eff}^2} \left|\frac{D_{23}}{V_{cb}}\right|^2 \left(\frac{x_e}{4}\right)^2 c_d\sqrt{ c_e} \, , 
\end{equation}
and, in the Extended Model,
\begin{equation}
C_{\mu-e}^{\rm EM}=-\frac{1}{M_{\rm eff}^2} \left(\frac{s_{q2}}{s_{q3}} \frac{s_{l2}}{s_{l3}}\right)^2  \left(D^{L*}_{21} E^{L}_{22})(E^{L*}_{21} D^{L}_{21} \right)
\approx \frac{8.0\cdot 10^{-7} }{M_{\rm eff}^2} \left(\frac{y_q}{3} \frac{y_e}{3}\right)^2  c_d\sqrt{c_e} \, .
\end{equation}

\subsection{$B\rightarrow K \tau\mu$}

The effective Lagrangian relevant to $B^+\rightarrow K^+\tau^+\mu^-$ is
\begin{equation}
\mathcal{L}=C_{s\rightarrow b\mu\bar{\tau}} (\bar{b}_L\gamma_\mu \tau_L)(\bar{\mu}_L\gamma_\mu s_L) + {\rm h.c.} 
\end{equation}
where, in the Minimal Model,
\begin{equation}
C_{s\rightarrow b\mu\bar{\tau}}^{\rm MM}=-\frac{1}{M_{\rm eff}^2}  \left(D^{L*}_{33} E^{L}_{33})(E^{L*}_{32} D^{L}_{32} \right)
\approx \frac{6.8\cdot 10^{-3}}{M_{\rm eff}^2}  \left(\frac{D_{23}}{V_{cb}}\right)\left(\frac{x_e}{4}\right) \, , 
\end{equation}
and, in the Extended Model,
\begin{equation}
C_{s\rightarrow b\mu\bar{\tau}}^{\rm EM}=-\frac{1}{M_{\rm eff}^2}  \left(\frac{s_{q2}}{s_{q3}} \frac{s_{l2}}{s_{l3}}\right) 
\approx \frac{1.5\cdot 10^{-2}}{M_{\rm eff}^2}  \left(\frac{y_q}{3} \frac{y_e}{3}\right) \, .
\end{equation}
Similarly for $B^+\rightarrow K^+\tau^-\mu^+$ it is

\begin{equation}
\mathcal{L}=C_{s\rightarrow b\bar{\mu}\tau} (\bar{b}_L\gamma_\mu \mu_L)(\bar{\tau}_L\gamma_\mu s_L) +{\rm h.c.} 
\end{equation}
where, in the Minimal Model,
\begin{equation}
C_{s\rightarrow b\bar{\mu}\tau}^{\rm MM}=-\frac{1}{M_{\rm eff}^2} \left(D^{L*}_{33} E^{L}_{32})(E^{L*}_{33} D^{L}_{32} \right)
\approx \frac{6.8\cdot 10^{-3}}{M_{\rm eff}^2}  \left(\frac{D_{23}}{V_{cb}}\right)\left(\frac{x_e}{4}\right) \, , 
\end{equation}
and, in the Extended Model,
\begin{equation}
C_{s\rightarrow b\bar{\mu}\tau}^{\rm EM}=-\frac{1}{M_{\rm eff}^2} \left(\frac{s_{ql}}{c_{ql}}\right)^2\left(\frac{s_{q2}}{s_{q3}} \frac{s_{l2}}{s_{l3}}\right) \approx \frac{{6.8\cdot 10^{-3}}}{M_{\rm eff}^2}  \left(\frac{s_{ql}}{c_{ql}}\frac{y_e}{2}\right) \left(\frac{s_{ql}}{c_{ql}}\frac{y_q}{2} \right)  \, .
\end{equation}

\subsection{$\tau\rightarrow \mu\gamma$}

The $\tau\rightarrow \mu\gamma$ amplitude receives from the leptoquark exchange a one-loop contribution, which depends on the leptoquark interactions with the light fermions, eq.~(\ref{Lint_light}), on its minimal gauge invariant interactions with the hypercharge field and on the interaction
\begin{equation}
\Delta \mathcal{L}=-i g' \frac{2}{3} k_Y V_\mu^+ V_\nu B^{\mu\nu}.
\end{equation}
In terms of the effective Lagrangian
\begin{equation}
\mathcal{L} = C_{\tau\rightarrow \mu\gamma} e m_\tau (\bar{\mu}_L \sigma_{\mu\nu}\tau_R) F_{\mu\nu} + {\rm h.c.}
\end{equation}
the coefficient $C_{\tau\rightarrow \mu\gamma}$ in the Minimal Model can be written as 
\begin{align}
C_{\tau\rightarrow \mu\gamma}^{\rm MM} & = \frac{1}{M_{\rm eff}^2} \frac{A}{32\pi^2}  E^{L*}_{32} E^{L}_{33}
\approx \frac{5.2 \cdot 10^{-4}}{M_{\rm eff}^2} A \left(\frac{x_e}{4}\right) \, , 
& 
A & = (1-k_Y) \left( \log{\frac{\Lambda^2}{M_V^2}} + \frac{3}{2} \right)  - 1 \, .
\label{taumuMM}
\end{align}
In a similar way in the Extended Model
\begin{equation}
C_{\tau\rightarrow \mu\gamma}^{\rm EM} = \frac{1}{M_{\rm eff}^2} \frac{A}{32\pi^2}  \left(\frac{s_{ql}}{c_{ql}}\right)\left(\frac{s_{l2}}{s_{l3}}\right)
\approx \frac{2.6 \cdot 10^{-4}}{M_{\rm eff}^2} A \left(\frac{s_{ql}}{c_{ql}}\frac{y_e}{2}\right) \, .
\label{taumuEM}
\end{equation}
In eqs.~\eqref{taumuMM}, \eqref{taumuEM} we have  considered only the exchange of light down-quarks in the loop, as the  exchange of their partners depends on unknown heavy masses, which can be comparable to $M_V$.

\subsection{$\mu\rightarrow e\gamma$}

In terms of the effective Lagrangian
\begin{equation}
\mathcal{L} = C_{\mu\rightarrow e\gamma} e m_\mu (\bar{e}_L\sigma_{\mu\nu}\mu_R) F_{\mu\nu} + {\rm h.c.}
\end{equation}
the coefficient $C_{\mu\rightarrow e\gamma}$ in the Minimal Model is
\begin{equation}
C_{\mu\rightarrow e\gamma}^{\rm MM} = \frac{1}{M_{\rm eff}^2} \frac{A}{32\pi^2}  E^{L}_{32}E^{L*}_{31}
\approx \frac{5.9 \cdot 10^{-6}}{M_{\rm eff}^2} A \left(\frac{x_e}{4}\right)^2\sqrt{c_e} \, , 
\end{equation}
whereas in the Extended Model
\begin{equation}
C_{\mu\rightarrow e\gamma}^{\rm EM} = \frac{1}{M_{\rm eff}^2} \frac{A}{32\pi^2}   \left(\frac{s_{ql}}{c_{ql}}\right)^2\left(\frac{s_{l2}}{s_{l3}}\right)^2
E^{L*}_{21}
\approx \frac{1.5 \cdot 10^{-6}}{M_{\rm eff}^2} A \left(\frac{s_{ql}}{c_{ql}}\frac{y_e}{2}\right)^2 \sqrt{c_e} \, .
\end{equation}

\subsection{$\Delta B_{s,d}=2$}

The effective Lagrangian for $\Delta B=2$ transitions is generated by quadratically divergent loop effects. 
In the $\Delta B_s=2$ case
\begin{equation}
\mathcal{L} = C_{\Delta B_s=2} (\bar{s}_L\gamma_\mu b_L)^2 + {\rm h.c.} 
\end{equation}
where, in the Minimal Model,
\begin{equation}
C_{\Delta B_s=2}^{\rm MM}=-\frac{1}{M_{\rm eff}^4} \frac{\Lambda^2}{128\pi^2}\left(D^{L*}_{32}  D^{L}_{33} \right)^2
\approx  \frac{3.3 \cdot 10^{-5}}{M_{\rm eff}^2} \left(\frac{\TeV}{M_{\rm eff}} \right)^2 \left(\frac{\Lambda}{5 ~\TeV}\right)^2 \left(\frac{D_{23}}{V_{cb}}\right)^2 \, , 
\end{equation}
and, in the Extended Model,
\begin{equation}
C_{\Delta B_s=2}^{\rm EM}=-\frac{1}{M_{\rm eff}^4} \frac{\Lambda^2}{128\pi^2}\left(\frac{s_{ql}}{c_{ql}}\right)^2\left(\frac{s_{q2}}{s_{q3}}\right)^2
\approx \frac{1.3 \cdot 10^{-4}}{M_{\rm eff}^2} \left(\frac{\TeV}{M_{\rm eff}}\right)^2 \left(\frac{\Lambda}{5 ~\TeV}\right)^2 \left(\frac{s_{ql}}{c_{ql}}\frac{y_q}{2}\right)^2 \, .
\end{equation}
Similarly, in the $\Delta B_d=2$ case
\begin{equation}
\mathcal{L} = C_{\Delta B_d=2} (\bar{d}_L\gamma_\mu b_L)^2 + {\rm h.c.}
\end{equation}
where, in the Minimal Model,
\begin{equation}
C_{\Delta B_d=2}^{\rm MM}=-\frac{1}{M_{\rm eff}^4} \frac{\Lambda^2}{128\pi^2}\left(D^{L*}_{31} D^{L}_{33} \right)^2
\approx  \frac{1.7\cdot 10^{-6}}{M_{\rm eff}^2} \left(\frac{\TeV}{M_{\rm eff}}\right)^2 \left(\frac{\Lambda}{5 ~\TeV}\right)^2 \left(\frac{D_{23}}{V_{cb}}\right)^2 c_d \, , 
\end{equation}
and, in the Extended Model,
\begin{equation}
C_{\Delta B_d=2}^{\rm EM}=-\frac{1}{M_{\rm eff}^4} \frac{\Lambda^2}{128\pi^2}\left(\frac{s_{ql}}{c_{ql}}\right)^2\left(\frac{s_{q2}}{s_{q3}}\right)^2\frac{m_d}{m_s} c_d
\approx \frac{6.8 \cdot 10^{-6}}{M_{\rm eff}^2} \left(\frac{\TeV}{M_{\rm eff}}\right)^2 \left(\frac{\Lambda}{5 ~\TeV}\right)^2 \left(\frac{s_{ql}}{c_{ql}}\frac{y_q}{2}\right)^2 c_d \, .
\end{equation}
The current bounds on $C_{\Delta B_{d,s}=2}$~\cite{Silvestrini:2018dos} depend on their phases and are weakest for approximately real Wilson coefficients, giving the bounds that we quote in Table~\ref{DeltaB=2}.

\section{Summary and Outlook}
\label{summary}

 \begin{table}[t]
$$\begin{array}{c|c|c|c}
&M_{\rm eff}^2 C^{\rm MM}&M_{\rm eff}^2 C^{\rm EM}&\TeV^2 C\\ \hline
K_L\rightarrow \mu e&1.0 \cdot 10^{-6} \left|\frac{D_{23}}{V_{cb}}\right|^2 \left(\frac{x_e}{4}\right)^2\sqrt{c_d c_e}&5.1\cdot10^{-6} \left(\frac{y_q}{3} \frac{y_e}{3}\right)^2\sqrt{c_d c_e}&1.0 \cdot 10^{-5}\\ \hline
\mu N\rightarrow e N &1.6 \cdot 10^{-7} \left|\frac{D_{23}}{V_{cb}}\right|^2 \left(\frac{x_e}{4}\right)^2 c_d\sqrt{ c_e}&8.0 \cdot 10^{-7} \left(\frac{y_q}{3} \frac{y_e}{3}\right)^2 c_d\sqrt{c_e}&1.7 \cdot 10^{-6}\\ \hline
B^+\rightarrow K^+ \tau^+\mu^-&6.8\cdot 10^{-3} \left(\frac{D_{23}}{V_{cb}} \right)\left(\frac{x_e}{4}\right)&1.5\cdot 10^{-2}  \left(\frac{y_q}{3} \frac{y_e}{3}\right)&6.2 \cdot 10^{-2}\\ \hline
B^+\rightarrow K^+ \tau^-\mu^+&6.8\cdot 10^{-3} \left(\frac{D_{23}}{V_{cb}} \right)\left(\frac{x_e}{4}\right)&6.8\cdot 10^{-3} \left(\frac{s_{ql}}{c_{ql}}\frac{y_e}{2}\right) \left(\frac{s_{ql}}{c_{ql}}\frac{y_q}{2}\right) &7.9\cdot 10^{-2}\\ \hline
\tau\rightarrow \mu\gamma&5.2 \cdot 10^{-4} A \left(\frac{x_e}{4}\right)&2.6 \cdot 10^{-4} A \left(\frac{s_{ql}}{c_{ql}}\frac{y_e}{2}\right)&8.6 \cdot 10^{-4}\\ \hline
 \mu\rightarrow e\gamma&5.9 \cdot 10^{-6} A \left(\frac{x_e}{4}\right)^2\sqrt{c_e}&1.5 \cdot 10^{-6} A \left(\frac{s_{ql}}{c_{ql}}\frac{y_e}{2}\right)^2 \sqrt{c_e}&1.1 \cdot 10^{-6}\\ \hline
\end{array}$$
\caption{Predictions for the coefficients in the relevant effective Lagrangians, as defined in the text, compared with the current bounds  in the last column. }
\label{Kmue}
\end{table}

 \begin{table}[t]
$$\begin{array}{c|c|c|c}
&(M_{\rm eff}^4/\TeV^2)(5~\TeV/\Lambda)^2 C^{\rm MM}&M_{\rm eff}^4/\TeV^2)(5~\TeV/\Lambda)^2 C^{\rm EM}&\TeV^2 C\\ \hline
\Delta B_s=2&3.3 \cdot 10^{-5} \left(\frac{D_{23}}{V_{cb}}\right)^2 &1.3 \cdot 10^{-4} \left(\frac{s_{ql}}{c_{ql}}\frac{y_q}{2}\right)^2&2.2\cdot 10^{-5}\\ \hline
\Delta B_d=2&1.7\cdot 10^{-6} \left(\frac{D_{23}}{V_{cb}}\right)^2 c_d &6.8 \cdot 10^{-6} \left(\frac{s_{ql}}{c_{ql}}\frac{y_q}{2}\right)^2 c_d&1.0 \cdot 10^{-6}\\ \hline
\end{array}$$
\caption{Predictions for the coefficients in the relevant effective Lagrangians, as defined in the text, compared with the current bounds  in the last column~\cite{Silvestrini:2018dos}. }
\label{DeltaB=2}
\end{table}

The apparently emerging anomalies in the semi-leptonic decays of the B-mesons~\cite{Lees:2013uzd,Aaij:2015yra,Hirose:2016wfn,Aaij:2014ora,Aaij:2017deq,Aaij:2017vbb} have triggered a great interest both in the theoretical as in the experimental community. This is justified by the potential significance of these results and, even more importantly,  by the foreseen power of future data to prove, or disprove,  the reality of these anomalies with great precision~\cite{Bediaga:2018lhg,Kou:2018nap,Cerri:2018ypt}. To us a more specific reason comes from the involvement of third generation particles, three out of four particles  in the CC case. On one side this goes well with the relative isolation of the third generation particles from the first two, both in the spectrum and in the CKM angles, making the third generation particles special.  On the other side this very feature allows to conceive  detectable deviations from the SM  without conflicting with the extended body  of already existing data in the flavour sector. In both cases an approximate $U(2)$-symmetry may come into play,  that acts on the first two generations as doublets and the third generation particles as singlets.

This point of view, also considering the still  evolving character of the data on the anomalies, has motivated us to consider  models based on $U(2)$ that can catch some features of the SM parameters in the flavour sector and that, at the same time,  may lead to violations of LFU in $b$-decays at an observable level in foreseen experiments. To this end, at least as an example, we attribute the violations of LFU to the exchange of a vector leptoquark, $V_\mu^a$, singlet under $SU2)$ and carrying charge $2/3$. We end up with two models based on specific charges under $U(2)$ of the standard fermions $f$ and of the mediator heavy fermions $F$, which both give rise to the  predictions of the CKM angles described in Section~\ref{predictions}.

The expected range for the observable violations of LFU in $b$-decays is shown in Figs.~\ref{fig:D32<U32} and \ref{fig:anomalies}. Fig.~\ref{fig:D32<U32} refers to the Minimal Model (MM), so called because of the simple transformation properties under a single $U(2)$-symmetry of both the light and the heavy fermions. As such, the MM
only involves, other than the effective scale $M_{\rm eff}$, three ${\cal O}(1)$ parameters, $D_{23}/V_{cb}, E_{23}/V_{cb}$ and $\tan\theta_e$. The Extended Model (EM) involves several ${\cal O}(1)$parameters, some of which are assumed dominant when Fig.~\ref{fig:anomalies} is drawn. While the size  of the expected anomalies are significantly different in the two cases, based on existing forecasts we think that the ranges in the two figures will be explored  in foreseen experiments. Note in particular that in the MM the predicted values of the anomalies, Fig.~\ref{fig:D32<U32}, are below the central values of the current data, eq.~\eqref{CCdata}, which are, however, still evolving.
More specific conclusions drawn from these figures are: 
\begin{itemize}
\item In the MM, $M_{\rm eff}$ can be higher than the range shown (i.e. $M_{\rm eff}< 1.5 \, \TeV$,  which is expected to be fully explorable at LHC with $3 ~{\rm ab}^{-1}$ of integrated luminosity~\cite{Schmaltz:2018nls,Baker:2019sli}) only at the price of making the violation of LFU in the CC case invisible. 

\item In the EM, $M_{\rm eff}$ can be higher than  $1.5 \, \TeV$ with violations of LFU  still observable  both in the CC and NC cases with reasonable ${\cal O}(1)$ parameters. 
\end{itemize}
A  relatively precise description in both models of the first two generations makes it possible to predict  a number of flavour-violating observables in a restricted range.  For some of these observables, the corresponding ranges are summarised in Table~\ref{Kmue} and compared with the bounds from current experiments for the coefficients of the relevant effective operators. The ${\cal O}(1)$ parameters occurring in these predictions, all shown in the Table, are normalised to their most likely values, depending on the internal consistency of the picture in both models. The constraint from $\mu\rightarrow e \gamma$ appears particularly significant for the MM.

Needless to say that the UV completion of a vector leptoquark exchange is non-trivial~\cite{Barbieri:2016las,Diaz:2017lit,DiLuzio:2017vat,Cline:2017aed,Assad:2017iib,Calibbi:2017qbu,Bordone:2017bld,Barbieri:2017tuq, Blanke:2018sro,Bordone:2018nbg,DiLuzio:2018zxy,Cornella:2019hct} and, no doubt, will be required in case the anomalies will be confirmed at some level. This will bring in a number of new effects as of low-energy relevant effective operators. At the same time this will allow a fully meaningful treatment of matching and RG-running effects, known to be  potentially significant~\cite{Feruglio:2016gvd,Feruglio:2017rjo,Crivellin:2018yvo}.
At this stage we have limited ourselves to show, with a  cutoff $\Lambda$, what is likely to be one of the most relevant, if not the most relevant, loop effect: $\Delta B=2$ transitions with  leptoquark exchanges. The corresponding results are summarised in Table~\ref{DeltaB=2}. The constraints appear severe for the EM, but  one should not forget, other than  possible extra contributions occurring in a proper UV completion, the assumed dominance of some parameters, as recalled above and in Section~\ref{epsilon}.
 
{\small
\subsubsection*{Acknowledgements}
We thank  Dario Buttazzo, Gino Isidori, Luca di Luzio, Marco Nardecchia and Luca Silvestrini for useful comments and discussions. We also thank the Galileo Galilei Institute for Theoretical Physics for the hospitality and the INFN for partial support during the completion of this work.}


\end{document}